\documentclass{aa}
\usepackage{multirow}
\usepackage{newtxtext,newtxmath}
\usepackage[T1]{fontenc}
\usepackage{ulem}
\DeclareRobustCommand{\VAN}[3]{#2}
\let\VANthebibliography\thebibliography
\def\thebibliography{\DeclareRobustCommand{\VAN}[3]{##3}\VANthebibliography}
\usepackage{graphicx}   
\usepackage{xcolor}
\usepackage{txfonts}
\usepackage[switch, modulo]{lineno}
\usepackage{longtable}

\newcommand{\bz}{\ensuremath{\langle B_z \rangle}}
\newcommand{\bs}{\ensuremath{\langle \vert B \vert \rangle}}

\newcommand{\teff}{\ensuremath{T_{\rm eff}}}

\begin{document}

   \title{Discovery of magnetic fields in five DC white dwarfs}
\titlerunning{Discovery of magnetic fields in five DC white dwarfs}   
    \author{ Andrei V. Berdyugin\inst{1}
                 \and 
           Vilppu Piirola\inst{1}
          \and
          Stefano Bagnulo\inst{2}
          \and
          John D. Landstreet\inst{2,3}
          \and
          Svetlana V. Berdyugina\inst{4,5,6}
          }
   \institute{Department of Physics and Astronomy, FI-20014 University of Turku, Finland;
              \email{andber@utu.fi}                                                            
         \and
             Armagh Observatory \& Planetarium, College Hill, Armagh BT61 9DG, UK;
         \and
         Department of Physics \& Astronomy, University of Western Ontario, London, Ontario N6A 3K7, Canada;
         \and 
           Leibniz-Institut f\"ur Sonnenphysik (KIS), Sch\"oneckstr 6, Freibirg, Germany;
       \and
       IRSOL Istituto Ricerche Solari “Aldo e Cele Dacc\`o", Faculty of Informatics, Universit\`a della Svizzera italiana, Via Patocci 57, Locarno, Switzerland;         
       \and
       Euler Institute, Faculty of Informatics, Universit\`a della Svizzera italiana, Via la Santa 1, 6962 Lugano, Switzerland
           }

  \date{Received October 7, 2022; accepted October 27, 2022}

\abstract{
About half of white dwarfs (WDs) evolve to the DC state as they cool; the others become DQ or (temporarily?) DZ WDs. The recent magnetic survey of the local 20\,pc volume has established a high frequency of magnetic fields among WDs older than 2--3\,Gyr, demonstrating that in low- and average-mass WDs, the effects of magnetism become more common as they age, and the fields on average become stronger. However, the available statistics of WDs older than about 5 Gyr do not clearly establish how fields evolve beyond this age. We are carrying out a survey to clarify the occurrence of magnetism in DC-type WDs in order to better understand this late evolution. We use broadband filter polarimetry, arguably the most efficient way to detect magnetic fields in featureless WDs via continuum circular polarization. Here we report the discovery of a magnetic field in five DC WDs (of 23 observed), almost doubling the total sample of known magnetic WDs belonging to the DC spectral class. 
}

 \keywords{White dwarfs -- Stars: magnetic fields -- polarization
              }

  \maketitle


\section{Introduction}
Single stars of $M \la 8 M_\odot$ evolve to become white dwarfs (WDs). The descendants of these single stars of intermediate mass provide most of the population of WDs, concentrated around the mean mass of $0.6 M_\odot$. A smaller fraction of current WDs were also formed from close binary systems. Some of these systems eventually  merged to form a single collapsed remnant, frequently of a significantly larger mass; others ended their nuclear lifetimes as double WD binaries. 

Once formed, the evolution of a WD is normally to cool slowly over several gigayears. Cooling is a fairly complex process even for single-star evolution, both to observe and to understand. Observationally, young hot WDs usually show strong spectra of H (DA WDs), He (DB WDs), or sometimes C (DQs). As they cool, spectral lines of the dominant elements H or He become weaker: He lines vanish at about 11000\,K, H lines around 5000\,K. In parallel with this general evolution, WDs may (temporarily) show lines of metals such as Mg, Si, Ca, and/or Fe (DZ, DAZ, and DZA stars), and some have spectra dominated by C. Below about 5000\,K, about one-quarter of WDs have very weak H$\alpha$, another quarter show spectral lines of metals (especially Ca\,{\sc ii}) or of C$_2$, and the remaining half show essentially featureless spectra  \citep[DC WDs;][Table 1]{BagLan21}. It appears that the dominant element(s) in the atmosphere can change as cooling occurs, for example due to gravitational diffusion, development of convection, and accretion of circumstellar planetary debris. 

One of the physical effects adding complexity to our efforts to understand WD evolution is that a significant fraction, about 20-25\%, of WDs in the local volume near the Sun \citep{BagLan21} possess detectable surface magnetic fields \citep[this high frequency was already suggested on the basis of literature reports of fields in nine WDs in the 13\,pc volume by ][]{Kawetal07}. The fields observed at the surface range in strength, measured by the mean surface field \bs, from tens of kG to hundreds of MG. Such fields can significantly affect WD evolution by altering or suppressing surface convection and internal shear, and by transferring angular momentum between internal layers or during accretion or mass loss \citep[see for example][]{Treetal15}. The fields may also introduce additional forces into envelope and atmosphere layers, altering their hydrostatic structure from that expected when magnetic effects are absent \citep{Lan87}.

For WDs formed by single-star evolution, which generates most of the large populations of WDs with masses around $0.6 M_\odot$, it has become clear that recently formed WDs (with cooling ages of less than, say, 1 Gyr) are very rarely detectably magnetic, and when they are magnetic, the fields are usually very weak \citep{BagLan22}. As WDs cool, fields begin to appear more frequently and usually become stronger. In WDs older than 3 or 4\,Gyr, megagauss-scale fields are not uncommon \citep{BagLan21}.

The observed evolution in magnetic field frequency and strength of normal-mass WDs for the first few gigayears of cooling may be understood as a slow emergence -- as a result of field relaxation to the stellar surface -- of the internal fields present in the degenerate cores of the WD precursors. An additional contribution to observed surface fields may be due to magnetic fields generated during cooling by a dynamo that acts during the period when the core of the WD is crystallising \citep{Iseetal17,Genetal18}. Beyond the end of crystallisation, the only identified evolution mechanisms are continued field relaxation and Ohmic decay. 

Observationally, however, after about 5 Gyr of WD cooling, we have very limited information with which to guide and confront theory. Only small survey samples constrain observed field evolution on cool WDs, such as DQ WDs, where C$_2$ bands show no polarization in strong fields \citep{Berdetal2007}. Particularly little is known about the magnetic fields of DC WDs, in which no spectral features are seen at all, leading to the questions of whether field strength begins to decay Ohmically and whether the frequency of surface fields continues to increase. Data that could help us answer these questions are very limited. For WDs within 20 pc of the Sun (the 20\,pc volume sample), \citet{BagLan21} showed that of 31 DC WDs, only 4 are magnetic white dwarfs (MWDs), and that only 4 of 24 WDs of any spectral class older than $\sim 6$\,Gyr are magnetic. These data are obviously too limited to clearly describe the evolution of fields in these old WDs. 

Previous surveys have provided almost no information about magnetic fields in DC WDs. Fields in such stars cannot be detected through the magnetic splitting of spectral lines. They can only be detected via the observation of continuum circular polarisation \citep[CCP;][]{Kempb}, a method of observation hardly employed since the 1970s \citep{Angetal81}. Remarkably, most CCP observations have led to the discovery of magnetic fields in stars that are not featureless but in which the magnetic field is strong enough to shift and broaden spectral lines in a such a way as to make their intensity spectra unrecognisable. Only seven featureless DC WDs are presently known to be magnetic. Five of them were discovered only in the last couple of years \citep{BagLan20,Beretal22}. Before these results, the only known magnetic DC stars were G195-19 and G111-49, discovered respectively by \citet{AngLan71-Second} and \citet{Putney95}.  

To improve our knowledge of the magnetic fields in the latest stages of stellar evolution, we have started a volume-limited survey of DC stars in the local 33\,pc volume, which is about 4.5 times larger than the previously explored 20\,pc volume and should have a correspondingly larger sample of DCs and DC MWDs. With this sample we expect to find enough DC MWDs to delineate the evolution of their magnetic fields, both in the WDs with He-rich atmospheres that become DCs as soon as their effective temperatures reach about 11\,000\,K (`young' DCs), and in DC WDs with \teff\ below about 5000\,K, with cooling ages of around 4\,Gyr or more (`old' DCs).

\section{Observations}

Almost all known MWDs have been discovered via the magnetic (Zeeman) splitting of spectral lines, observed in stellar flux spectra, or via the Zeeman polarisation of spectral features \citep{Feretal15}. Using these methods, fields of a few kG up to 1 GG can be reliably detected. However, these techniques cannot be used to measure fields in WDs that lack spectral lines. For such stars, it is necessary to rely on continuum polarisation, which \citet{Kempb} showed should occur in radiation from a magnetized emitter. The value of this effect was confirmed by the discovery of a very strong field in the bright WD Grw+70\,8247 = WD\,1900+705 through the detection of broadband circular polarisation (BBCP) by \citet{Kemetal70}. 

Broadband circular polarisation is a relatively weak effect. \citet{BagLan20} have estimated that a field of $\bz \sim 15$\,MG is required to produce BBCP of order 1\,\% in optical radiation from a cool WD. However, with a sensitive polarimeter, especially one with a very stable and well-established zero point, it is possible in principle to detect polarisation of $10^{-4}$ or less, corresponding to $ \sim 100$\,kG fields in `sufficiently bright' WDs. 

\begin{table*}
\caption{\label{Tab_Log_Std} Observing log of bright non-polarised standard stars and the highly polarised MWD WD\,1900+705. Polarisation values are given assuming as instrumental polarisation the values of 0.0121$\pm$0.0004\,\%, 0.0109$\pm$0.0005\,\%, and 0.0084$\pm$0.0004\,\% in the $B'$, $V'$, and $R'$ filters, respectively. For comparison, we report the polarisation values of WD\,1900+705 measured in our 2021 and 2022 runs.}
\begin{center}
\begin{tabular}{lrcccrr@{$\pm$}lr@{$\pm$}lr@{$\pm$}l}
\hline\hline
\multicolumn{1}{c}{STAR} & $G$ & DATE & UT & JD -- & Exp. & \multicolumn{6}{c}{$V/I$ (\%)} \\
                         &     &yyyy-mm-dd & hh:mm &2400000 &\multicolumn{1}{c}{(s)} &
\multicolumn{2}{c}{$B'$} &\multicolumn{2}{c}{$V'$} &\multicolumn{2}{c}{$R'$} \\
\hline
HD\,107146   & 6.9 & 2022-06-27 & 21:23 & 59758.391 & 1680 &$ 0.0005$& 0.0007 &$ -0.0006$& 0.0007 &$-0.0001$& 0.0006 \\
HD\,115043   & 6.7 & 2022-06-28 & 21:21 & 59759.390 & 1520 &$ 0.0005$& 0.0008 &$ -0.0001$& 0.0012 &$-0.0004$& 0.0004 \\
HD\,122652   & 7.0 & 2022-06-29 & 21:18 & 59760.387 & 1520 &$-0.0012$& 0.0009 &$ -0.0015$& 0.0015 &$-0.0019$& 0.0014 \\
HD\,122676   & 7.1 & 2022-06-30 & 21:17 & 59761.387 & 1520 &$ 0.0000$& 0.0011 &$  0.0003$& 0.0010 &$ 0.0018$& 0.0008 \\
HD\,124694   & 7.0 & 2022-07-01 & 21:16 & 59762.386 & 1520 &$-0.0012$& 0.0008 &$  0.0014$& 0.0011 &$ 0.0002$& 0.0007 \\
HD\,135891   & 6.9 & 2022-07-02 & 21:18 & 59763.387 & 1520 &$ 0.0014$& 0.0008 &$  0.0006$& 0.0010 &$-0.0004$& 0.0007 \\
HD\,117860   & 7.2 & 2022-07-03 & 21:16 & 59764.386 & 1520 &$-0.0004$& 0.0010 &$ -0.0002$& 0.0012 &$ 0.0005$& 0.0006 \\[2mm]
WD\,1900+705 &13.2 & 2021-07-02 & 22:22 & 59398.432 & 640 &$ 3.756 $& 0.016  &$  3.604 $& 0.016  &$ 3.827 $& 0.019  \\
             &     & 2022-07-02 & 00:25 & 59762.518 & &$ 3.789 $& 0.016  &$  3.602 $& 0.019  &$ 3.838 $& 0.018  \\
\hline
\end{tabular}
\end{center}
\end{table*}

To detect and measure broadband continuum polarisation, one uses either spectropolarimetry or filter polarimetry with broad, photometry-like filters. It is very difficult to establish the zero point with sufficient accuracy below polarisation levels of the order of $10^{-3}$ in spectropolarimetric measurements of the continuum \citep{Fosetal07,Sieetal14}; therefore, in DC WDs, only megagauss-scale fields can be detected in this way \citep{BagLan20}. In contrast, broadband filter polarimeters can be very stable, and instrumental polarisation can be calibrated at the $10^{-5}$ level, so detections with such instruments of fields of hundreds of kilogauss are in practice limited by the telescope aperture and WD brightness \citep{Beretal22}.

The search for magnetic fields of a fraction of 1 MG or stronger in DC WDs that is reported here was carried out with the DIPol-UF broadband filter polarimeter \citep{Piietal20} mounted on the 2.5 m Nordic Optical Telescope (NOT) at the Observatorio del Roque de los Muchachos on the island of La Palma, in the Canaries.  This instrument obtains simultaneous circular polarisation (normalized Stokes V/I) measurements in three filter bands isolated by dichroic mirrors. The passbands are centred at about 4450\,\AA\   (the $B'$ band), 5400\,\AA\ (the $V'$ band), and 6400\,\AA\ (the $R'$ band) with full widths at half maximum (FWHMs) of 1140, 750, and 960\,\AA\,, respectively. With this instrument on the NOT, we can detect a polarisation degree at the $3 \sigma$ level of $\sim 10^{-4}$ for the Gaia G-band magnitude $G \sim 12$, down to a degree of $\sim 10^{-3}$ at $G \sim 17$. This instrument and the filter system are discussed in more detail in our previous paper, which reports the results of the first part of our search for magnetic fields in DC WDs \citep{Beretal22}.

Here we report observations of 23 DC WDs and discovery of 5 new DC MWDs. We note that our survey includes nine young DCs of He-rich atmospheres with $11000 \la \teff \la 5000$\,K and 14 old WDs with $\teff \la 5000$\,K and ages $\tau \ga 4$\,Gyr. The stars are selected from available classifications and with help from \citet{Genetal21}. Our new observations were obtained between June 27 and July 5, 2022.

  \begin{figure}
  \centering
  \includegraphics[width=7cm]{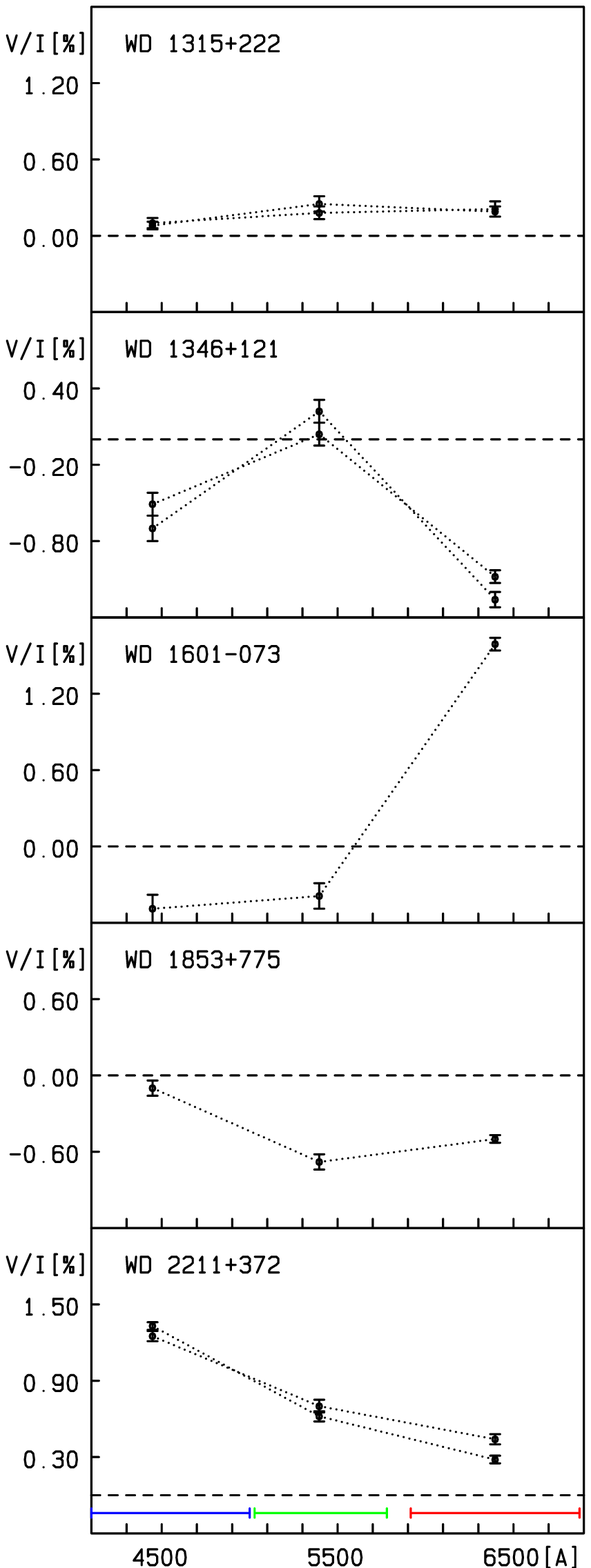}
  \caption{\label{Pcplot3} Wavelength dependence of circular polarisation detected for five targets ($>3\sigma$ confidence level). A wide variety of polarisation behaviour is observed. Horizontal bars in the bottom panel show the FWHM of the $B'V'R'$ filter passbands.}
  \end{figure}

\subsection{Instrumental polarisation and alignment of the polarimetric optics}

\begin{table*}
  \caption{\label{Tab_Programme} Programme stars and their main physical features. Star names in boldface identify WDs in which fields were discovered during the observations reported in this paper (see Table~\ref{Tab_Log_WD}). }
\begin{tabular}{llrrrcccl}
\hline\hline
\multicolumn{2}{c}{STAR}               &  $G$ &\multicolumn{1}{c}{$d$}& \teff & log $g$ &    $M$      & Age  & Atmosphere and ref. \\
               &                       &      &(pc)  & (K)   &  c.g.s. & ($M_\odot$) &(Gyr) & \\
\hline
WD\,0005+395    & LP 240-30            & 16.6 & 34.4 & 4680 & 6.77 & 0.08 & 1.40 & DC,  H Prob$_{\rm WD}=0.70$ (1,3)\\
WD\,0010+543    & LSR J0013+5437       & 18.0 & 32.3 & 4123 & 7.77 & 0.46 & 7.08 & DC,  (2, H assumed)  \\
WD\,0028+035    & PB 6002              & 16.1 & 27.8 & 6548 & 8.14 & 0.68 & 2.40 & DC,  (2, H assumed) \\
WD\,1251+366    & LP 267-311           & 17.2 & 28.5 & 4445 & 7.62 & 0.37 & 3.78 & DC,  He (1) \\
{\bf WD\,1315+222} & {\bf LP 378-956 } & 16.7 & 31.8 & 6235 & 8.21 & 0.71 & 3.61 & DCH, He (1) \\  
{\bf WD\,1346+121} & {\bf LP 498-66}   & 17.8 & 28.3 & 4150 & 7.88 & 0.50 & 6.58 & DCH, He (1) \\
WD\,1425+495    & CSO 649              & 16.7 & 33.9 & 6895 & 8.41 & 0.85 & 3.77 & DC,  (2, H assumed) \\
WD\,1427$-$238  & LP 857-45            & 17.4 & 32.6 & 4866 & 7.90 & 0.52 & 5.40 & DC,  (2, H assumed)\\
WD\,1434+437    & LP 221-217           & 17.2 & 27.2 & 4685 & 7.93 & 0.54 & 6.30 & DC,  H-He (1) \\
WD\,1533+469    & LP 176-60            & 17.8 & 30.8 & 4310 & 7.83 & 0.48 & 6.45 & DC?, H (1) \\
{\bf WD\,1601$-$073} & {\bf LP 684-16} & 17.9 & 26.9 & 4920 & 8.55 & 0.94 & 9.83 & DCH, (2, H assumed)\\
WD\,1612+092    & LSPM J1614+0906      & 17.2 & 27.9 & 4775 & 7.90 & 0.52 & 5.57 & DC,  H (1) \\
WD\,1702$-$016  & LP 626-29            & 17.3 & 28.3 & 4700 & 7.94 & 0.54 & 6.50 & DC,  (2, H assumed)\\
WD\,1737+798    & LP 24-66             & 16.9 & 26.8 & 5535 & 8.28 & 0.75 & 5.72 & DC,  He (1) \\
WD\,1746+450    & GD 366               & 15.5 & 29.9 & 9331 & 8.47 & 0.90 & 1.72 & DC,  (2, H assumed)\\
WD\,1800+508    & LP 139-38            & 17.4 & 31.0 & 4635 & 7.85 & 0.48 & 5.12 & DC,  He-H (1)\\
{\bf WD\,1853+775}    & {\bf LP 25-7}  & 17.0 & 30.5 & 4850 & 7.74 & 0.43 & 3.63 & DCH, He (1) \\
WD\,2058+550    & LSR J2059+5517       & 17.1 & 22.7 & 4415 & 7.93 & 0.53 & 7.15 & DC,  H-He (1) \\
WD\,2109$-$295  & EC 21096-2934        & 15.1 & 32.8 & 9260 & 7.98 & 0.57 & 0.78 & DC,  He-H (3)\\
WD\,2152$-$280  & LP930-61             & 16.3 & 23.5 & 5220 & 7.85 & 0.48 & 3.68 & DC,  He (1) \\
{\bf WD\,2211+372}    & {\bf LP 287-35}& 16.8 & 29.2 & 6345 & 8.47 & 0.88 & 4.56 & DC?H, He (1) \\
WD\,2215+368    & LP 287-39            & 16.8 & 20.3 & 4485 & 7.92 & 0.53 & 6.80 & DC,  H (1) \\
WD\,2311$-$068  & G 157-34             & 15.3 & 25.9 & 7360 & 7.97 & 0.56 & 1.31 & DC,  He (1) \\
\hline
\end{tabular}

\noindent
Key to references: 
1: \citet{Bloetal19};
2: \citet{Genetal21};
3: \citet{Beretal21}.
Where not found in these references, ages have been interpolated using the tables from \citet{Bedetal20}.
\end{table*}

During our observing run we obtained seven observations of seven different bright nearby stars, which are believed to have zero circular polarisation, to check for instrumental polarisation. These observations are reported in Table~\ref{Tab_Log_Std}. 

As in our previous run in July 2021, the high S/N measurements of non-polarised stars yield the instrumental polarisation to a precision better than $10^{-5}$. In the $B'V'R'$ bands, the values of Stokes V/I are 0.0121 $\pm$ 0.0004 \%, 0.0109 $\pm$ 0.0005 \%, and 0.0084 $\pm$ 0.0004 \%, respectively. These are very close to the values obtained in 2021. The instrumental polarisation was subtracted from the observed
polarisation of all targets, including the measurements of the standard stars reported in Table\,\ref{Tab_Log_Std}. 

In addition, we obtained one measurement of the well-known MWD WD\,1900+705, which appears to show a signal of circular polarisation that is nearly constant with time \citep[see e.g.][]{BagLan19a}. Our new measurement is compared in Table~\ref{Tab_Log_Std} to one of the same star that we made during the July 2021 run. The agreement is very satisfactory and demonstrates that we can obtain measurements that are precise at the 0.02\,\%\ level for a $G$ = $13.2$ star with about 10 minutes of exposure time. This shows that the alignment of our polarimetric optics is stable over a few years.

\subsection{Results}

\begin{table*}
\tabcolsep=0.14cm
  \caption{\label{Tab_Log_WD} Observing log of WDs. Detections are marked in boldface.  
  }

\begin{center}
\begin{tabular}{lcccrr@{$\pm$}lr@{$\pm$}lr@{$\pm$}l}
\hline\hline
\multicolumn{1}{c}{STAR} &    DATE  &  UT &JD --  & Exp. & \multicolumn{6}{c}{$V/I$ (\%)} \\
                         &yyyy-mm-dd&hh:mm&2400000&\multicolumn{1}{c}{(s)}  &
  \multicolumn{2}{c}{$B'$}&\multicolumn{2}{c}{$V'$} &\multicolumn{2}{c}{$R'$} \\
\hline
WD\,0005+395 & 2022-07-06 & 04:38 & 59766.693 & 3900 &${\rm -0.017}$&{\rm 0.063}&${\rm -0.082}$&{\rm  0.056}&${\rm  0.028}$&{\rm 0.044}\\ [1mm]
WD\,0010+543 & 2022-07-05 & 04:12 & 59765.675 & 7100 &${\rm -0.028}$&{\rm  0.140}&${\rm -0.011}$&{\rm   0.129}&${\rm 0.094}$&{\rm 0.067}\\ [1mm]
WD\,0028+035 & 2002-07-06 & 03:39 & 59766.652 & 3300 &${\rm -0.084}$&{\rm 0.045}&${\rm -0.051}$&{\rm  0.050}&${\rm -0.010}$&{\rm 0.058}\\ [1mm]
WD\,1251+366 & 2022-06-27 & 22:41 & 59758.445 & 5200 &${\rm -0.155}$&{\rm 0.065}&${\rm  0.006}$&{\rm  0.063}&${\rm  0.039}$&{\rm 0.038}\\ [1mm] 
{\bf WD\,1315+222} & 2022-06-28 & 22:25 & 59759.435 & 4200 &${\rm  0.104}$&{\rm 0.045}&${\bf  0.182}$&{\bf  0.052}&${\bf  0.215}$&{\bf 0.064}\\
                     & 2022-07-01 & 22:24 & 59762.434 & 4200 &${\rm  0.084}$&{\rm 0.040}&${\bf  0.253}$&{\bf  0.061}&${\bf  0.199}$&{\bf 0.044}\\ [1mm] 
{\bf WD\,1346+121} & 2022-07-02 & 22:43 & 59763.446 & 6500 &${\bf -0.508}$&{\bf 0.093}&${\rm  0.044}$&{\rm  0.091}&${\bf -1.074}$&{\bf 0.054}\\ 
                     & 2022-07-04 & 22:34 & 59765.44  & 6500 &${\bf -0.691}$&{\bf 0.109}&${\rm  0.222}$&{\rm  0.098}&${\bf -1.256}$&{\bf 0.062}\\ [1mm]
WD\,1425+495 & 2022-06-29 & 22:22 & 59760.432 & 4200 &${\rm  0.012}$&{\rm 0.045}&${\rm -0.002}$&{\rm  0.058}&${\rm 0.018}$&{\rm 0.038}\\ [1mm]
WD\,1427-238 & 2022-06-30 & 23:11 & 59761.466 & 5600 &${\rm -0.038}$&{\rm 0.114}&${\rm  0.102}$&{\rm  0.078}&${\rm  -0.044}$&{\rm 0.061}\\ [1mm]
WD\,1434+437 & 2022-06-30 & 00:04 & 59760.503 & 5200 &${\bf  0.231}$&{\bf 0.077}&${\rm -0.019}$&{\rm  0.067}&${\rm -0.019}$&{\rm 0.059}\\ [1mm]
WD\,1533+469 & 2022-07-01 & 01:02 & 59761.543 & 6600 &${\rm -0.065}$&{\rm 0.127}&${\rm -0.273}$&{\rm  0.110}&${\bf -0.195}$&{\bf 0.052}\\ 
                     & 2022-07-03 & 00:42 & 59763.529 & 6600 &${\rm  0.139}$&{\rm 0.098}&${\bf -0.300}$&{\bf  0.090}&${\rm -0.019}$&{\rm 0.042}\\ [1mm]
{\bf WD\,1601-073} & 2022-07-05 & 22:52 & 59766.453 & 6800 &${\bf -0.484}$&{\bf 0.111}&${\bf -0.386}$&{\bf 0.106}&${\bf 1.597}$&{\bf 0.054}\\ [1mm]
WD\,1612+092 & 2022-06-28 & 00:59 & 59758.541 & 5100 &${\rm  0.087}$&{\rm 0.069}&${\rm  0.033}$&{\rm  0.052}&${\rm  0.097}$&{\rm 0.046}\\ [1mm]
WD\,1702-016 & 2022-06-30 & 01:59 & 59760.582 & 5300 &${\rm  0.215}$&{\rm 0.091}&${\rm  0.087}$&{\rm  0.086}&${\rm  -0.104}$&{\rm 0.053}\\ [1mm]
WD\,1737+798 & 2022-06-29 & 01:34 & 59759.566 & 4500 &${\rm -0.014}$&{\rm 0.082}&${\rm -0.085}$&{\rm  0.093}&${\rm -0.071}$&{\rm 0.056}\\ [1mm]
WD\,1746+450 & 2022-06-28 & 02:12 & 59758.592 & 2600 &${\rm  0.012}$&{\rm 0.035}&${\rm -0.049}$&{\rm  0.032}&${\rm -0.074}$&{\rm 0.035}\\ [1mm]
WD\,1800+508 & 2022-07-05 & 00:59 & 59765.541 & 5600 &${\rm -0.157 }$&{\rm  0.073}&${\rm -0.047}$&{\rm  0.061}&${\rm  0.051}$&{\rm 0.053}\\ [1mm]
{\bf WD\,1853+775} & 2002-07-06 & 01:15 & 59766.552 & 4800 &${\rm -0.098}$&{\rm 0.066}&${\bf -0.680}$&{\bf  0.068}&${\bf -0.492}$ &{\bf 0.039}\\ [1mm]
WD\,2058+550 & 2022-07-02 & 04:15 & 59762.677 & 5000 &${\rm  0.068}$&{\rm 0.075}&${\rm -0.052}$&{\rm  0.070}&${\rm 0.064}$&{\rm 0.045}\\ [1mm]
WD\,2109-295 & 2022-07-01 & 03:51 & 59761.660 & 2200 &${\rm  0.011}$&{\rm 0.020}&${\rm  0.036}$&{\rm  0.034}&${\rm  -0.039}$&{\rm 0.037}\\ [1mm]
WD\,2152-280 & 2022-06-28 & 03:59 & 59758.666 & 3600 &${\rm  0.007}$&{\rm 0.040}&${\rm -0.055}$&{\rm  0.041}&${\rm -0.043}$&{\rm 0.022}\\ [1mm]
{\bf WD\,2211+372} & 2022-07-02 & 02:10 & 59762.590 & 4400 &${\bf  1.254}$&{\bf 0.041}&${\bf  0.703}$&{\bf  0.054}&${\bf  0.446}$&{\bf 0.044}\\     
                     & 2022-07-03 & 04:24 & 59763.683 & 4400 &${\bf  1.333}$&{\bf 0.038}&${\bf  0.623}$&{\bf  0.044}&${\bf  0.285}$&{\bf 0.040}\\ [1mm]
WD\,2215+368 & 2022-06-30 & 04:12 & 59760.675 & 4400 &${\rm  0.187}$&{\rm 0.083}&${\rm  0.133}$&{\rm  0.068}&${\rm  0.101}$&{\rm 0.044}\\ [1mm]
WD\,2311-068 & 2022-06-29 & 04:38 & 59759.693 & 2400 &${\rm -0.011}$&{\rm 0.028}&${\rm  0.011}$&{\rm  0.039}&${\rm  0.032}$&{\rm 0.032}\\ [1mm]
\hline
\end{tabular}
\end{center}
\end{table*}

The WDs observed during our 2022 June-July run are listed in Table\,\ref{Tab_Programme}, with their $G$ magnitudes, distances, physical parameters, cooling ages, and some comments. Physical parameters were obtained from various studies, cited in the table's notes; cooling ages are interpolated from the online cooling data provided by the Montreal group \citep{Bedetal20}. 

The observations are described in the log in Table\,\ref{Tab_Log_WD}, which gives dates, integration times, and the polarisation data in the three filter bands for each WD observation. We list measured BBCP values in boldface if non-zero polarisation is detected at above the $3 \sigma$ level. We consider that real polarisation has been detected if a consistent picture of detection is found across the bands, and we highlight star names of WDs in which polarisation is convincingly detected in boldface in Tables\,\ref{Tab_Programme} and \ref{Tab_Log_WD}. Of the 23 stars observed, BBCP has been definitely detected in 5 WDs. The data for these stars are plotted in Fig.\,\ref{Pcplot3}. 

We observed three of the five WDs in which fields were detected in order to fully confirm the weak field detections and to check for possible variability. No variability is detected with confidence. There are in addition two further WDs, WD 1434+437 and WD 1533+469, in which marginal polarisation detections have been obtained; these WDs await further observation to confirm (or not) the fields that may have been detected. However, a single pair of measurements does not probe all the possible timescales of variation; in particular, our measurements require integration of the order of one hour and so cannot probe all the rotation periods that might result from the formation of a MWD from a close binary. 

With these new discoveries, we almost double the number of DC WDs in which magnetic fields have been detected. One of the new MWDs discovered, LP\,684-16 = WD\,1601--073, is quite massive compared to most of the rest of the DC WDs observed.\ Therefore, because of its relatively small radius, it has cooled quite slowly, reaching only $\teff$ = 4920\,K, but has a computed cooling time of 9.8\,Gyr. It is probably the oldest magnetic WD of any spectral type discovered so far. For comparison, according to the parameters listed by \citet{BagLan21}, the oldest MWD in the 20 pc volume, in which such old MWDs are most likely to be discovered, is WD\,1008+290 = LHS\,2229. It is a DQpec star with an age of about 7.9\,Gyr, almost 2\,Gyr younger than LP\,684-16. 

We note that no really large polarisation signals, such as that exhibited by WD\,1900+705 (see Table\,\ref{Tab_Log_Std}), are found. However, the observed level of polarisation in three of the five definite detections reaches the range 1.2 to 1.6\%, so some of the fields detected are probably quite strong.

\section{Discussion and conclusions}
We continue to detect MWDs in roughly one-fifth of the DC sample observed. Considering that only relatively strong fields can be detected in featureless stars, our results suggest that the frequency of the occurrence of magnetic fields in older WDs may be as high as 25 or even 30\,\%, consistent with the frequency suggested by \citet{BagLan21}. 

Polarisation levels in the seven DC MWDs discovered by \cite{Beretal22} and in this paper range from about 0.1 to 1.6\,\%.  Using the order-of-magnitude estimator of  \citet{BagLan20} of a longitudinal field of 15\,MG, which leads to BBCP of the order of 1\%, inferred fields \bz\ are thus estimated to lie between perhaps 1 and 30\,MG. From this result, the fields \bs\ that we detect likely lie in the range of roughly 3 to 200\,MG.

Some of the fields produce a polarisation with the same sign in the three filter bands, while in other stars we detect a polarisation that reverses sign between one filter band and another (Fig. 1). Similar behaviour was found in our earlier survey data \citep{Beretal22} as well as in other strongly magnetic old WDs \citep{AngLan71-Second,Angetal74b,Angetal75,Putney95}. 

We carried out a second observation for three of our five new discoveries and for one suspected candidate. The confirming observations were obtained between one and three days after the discovery observations. For each of these four stars, the repeated observation confirms the presence of the magnetic field first detected, except for one $R'$ observation of WD\,1533+469. In no case do we detect clearly significant variability; we note, however, that our repeated measurements probe only a very limited range of timescales. 

We find that, in practice, a magnetic field can be reliably detected from BBCP at the polarization levels of approximately 0.2\%\ with our broadband filter polarimeter on a 2.5 m telescope in a DC WD of $G \sim 17$. We would gain about a factor of 3 in precision, or a 2.5 magnitude increase in limiting magnitude, by going to an 8 m telescope. 

To summarise the current statistical situation, we combine the results from the 20\,pc volume magnetic field survey \citep{BagLan21}, our exploratory observing run \citep{Beretal22}, and this work. We have collected literature data on and surveyed 30 young DC stars, of which 7 have been found to host magnetic fields, and 43 old DCs, of which 6 are magnetic. 

As more detections of DC MWDs are made, especially in the context of volume-limited surveys such as ours, comparisons with magnetic data for other types of WDs will at first be hampered by our current inability to assign precise field strength values to magnetic DC WDs. However, a first comparison can be made between the overall level of polarisation observed in young and old DC MWDs, which may be taken as an indicator of the evolution of the overall field strength with cooling age between these two populations. In the currently small sample, it appears that larger polarisation levels (say, above Stokes V/I $\ga 1$\,\%) appear to be at least as common in the old DC MWD population as in the young group. There is no clear signal of Ohmic field strength decay. However, the available sample is still very small, and as our sample increases we can hope to obtain a statistically more significant constraint and carry out modelling to provide more accurate field strength estimates corresponding to observed polarisation. Such surveys need to be carried on until a clearer statistical view of the magnetism in DC WDs is obtained.

\begin{acknowledgements}
Based on observations made with the Nordic Optical Telescope, owned in collaboration by the University of Turku and Aarhus University, and operated jointly by Aarhus University, the University of Turku and the University of Oslo, representing Denmark, Finland and Norway, the University of Iceland and Stockholm University at the Observatorio del Roque de los Muchachos, La Palma, Spain, of the Instituto de Astrofisica de Canarias. DIPol-UF is a joint effort between University of Turku (Finland) and Leibniz Institute for Solar Physics (Germany). We  acknowledge  support from the Magnus Ehrnrooth foundation and the ERC Advanced Grant HotMol  ERC-2011-AdG-291659.
JDL acknowledges the financial support of the Natural Sciences and Engineering Research Council of Canada (NSERC), funding reference number 6377-2016.
\end{acknowledgements}
\section{Data Availability}
All raw data and calibrations are available on request from the authors. 

\bibliographystyle{aa}
\bibliography{sbabib} 

\end{document}